\documentclass[a4paper,11pt]{amsart}
\usepackage{amssymb,amsmath,}
\newtheorem{theorem}{Theorem}[section]
\newtheorem{corollary}[theorem]{Corollary}

\newtheorem{prop}[theorem]{Proposition}
\theoremstyle{definition}
\newtheorem{definition}[theorem]{Definition}
\theoremstyle{remark}
\newtheorem{rem}[theorem]{Remark}
\numberwithin{equation}{section}
\begin{document}

\title{Gibbs States on Random Configurations}
\author{Alexei Daletskii}%
\address{Department of Mathematics, University of York, York YO10 5DD,
UK}
\email{alex.daletskii@york.ac.uk}%
\author{Yuri Kondratiev}
\address{Fakult\"{a}t f\"{u}r Mathematik, Universit\"{a}t Bielefeld,
D-33501 Bielefeld, Germany} \email{kondrat@math.uni-bielefeld.de}
\author{Yuri Kozitsky}
\address{Instytut Matematyki, Uniwersytet Marii Curie-Sklodowskiej,
20-031 Lublin, Poland } \email{jkozi@hektor.umcs.lublin.pl}
\author{Tanja Pasurek}
\address{Fakult\"{a}t f\"{u}r Mathematik, Universit\"{a}t Bielefeld,
D-33501 Bielefeld, Germany} \email{tpasurek@math.uni-bielefeld.de}

\begin{abstract}
Gibbs states of a spin system with the single-spin space
$S=\mathbb{R}^{m}$ and unbounded pair interactions is studied. The
spins are attached to the points of a realization $\gamma $ of a
random point process in $\mathbb{R}^{n}$. Under certain conditions
on the model parameters we prove that,  for almost all $\gamma $,
the set $\mathcal{G}(S^\gamma)$ of all Gibbs states is nonempty and
its elements have support properties, explicitly described in the
paper. We also show the existence of measurable selections $\gamma
\mapsto \nu _{\gamma }\in \mathcal{G}(S^\gamma)$ (random Gibbs
measures) and derive the corresponding averaged moment estimates.
\end{abstract}

\maketitle

\section{Introduction}

The aim of this paper is to study Gibbs sates (states of thermal
equilibrium) of the following system of interacting particles. The
underlying set is a countable collection  of point particles
chaotically distributed over a Euclidean space $X$, modeled by a
random point process in $X$. Each particle $x$ in the collection
possesses an internal structure described by a mark (spin) $\sigma
(x)$ taking values in a {\em single-spin space} $S_x$ and
characterized by a {\em single-spin measure} $\chi_x$. The system as
a whole is characterized by the law of the underlying point process
$\mu$,  by the spin-spin pair interaction dependent on the location
of the particles, and by the family of single-spin measures
$\{\chi_x\}_{x\in \mathbb{R}^d}$. For a fixed realization of the
point process $\gamma$, a Gibbs state is a probability measure on
the product space $S^{\gamma }=\prod_{x\in \gamma }S_{x}$
constructed in the following way. First we equip $S^{\gamma }$ with
the usual product topology and the corresponding Borel $\sigma
$-algebra $\mathcal{B}(S^{\gamma })$. Then we introduce the set
$\mathcal{P}(S^\gamma)$ of all probability measures on $(S^\gamma,
\mathcal{B}(S^\gamma))$. If the pair interaction is absent, i.e.,
the spins are independent, the unique Gibbs state is just the
product $\bigotimes_{x\in \gamma }\chi _{x}\in
\mathcal{P}(S^\gamma)$ of the single-spin measures. The states of
the system with interacting spins are constructed as perturbations
of the mentioned product measure  by the ``densities"
\begin{equation}
\exp \left( -\sum_{\{x,y\}\subset \mathcal{\gamma }}W_{xy}(\sigma
(x),\sigma (y))\right) ,  \label{D1}
\end{equation}%
where $W_{xy}:S\times S\rightarrow \mathbb{R}$ are measurable
functions -- {\em interaction potentials}. Clearly, (\ref{D1}) is
just a heuristic expression -- the rigorous definition is based on
the Gibbs specification constructed by means of the potentials
$W_{xy}$. Then the Gibbs states $\nu_\gamma$ are defined as elements
of $\mathcal{P}(S^\gamma)$ that solve  the Dobrushin-Lanford-Ruelle
(DLR) equation corresponding  to the Gibbs specification (see e.g.
\cite{Geor,Pre} and Introduction in \cite{KKP2}).

If the underlying set is fixed and reasonably regular, the only
problem which one faces in constructing Gibbs states of models with
interactions as in (\ref{D1}) is the possible unboundedness of the
potentials $W_{xy}$. Works in this direction were originated in
seminal papers \cite{LP,Ru70} where the underlying set is  a cubic
lattice\textit{\ }$\mathbb{Z}^{d}$ and the potentials are unbounded
functions on $\mathbb{R}\times \mathbb{R}$, see also \cite{KP} for a
more recent results. The case where the underlying set is a fixed
unbounded degree graph was studied in \cite{KKP1}.

In the present paper, we study Gibbs states of a spin system of this
kind with the underlying set chosen at random from the collection of
all locally finite subsets of a Euclidean space $X=\mathbb{R}^{n}$,
$n\geq 1$ (called 'simple configurations' in $X$); that is, from the
set
\begin{equation}
\Gamma (X)=\left\{ \gamma \subset X:\ N\left( \gamma \cap \Lambda
\right) <\infty ,\ \Lambda \in \mathcal{B}_{0}(X)\right\} ,
\label{gamma}
\end{equation}%
where $N(A)$ stands for  the cardinality of $A$ and
$\mathcal{B}_{0}(X)$ is the collection of all compact subsets of
$X$. The set $\Gamma (X)$ is endowed with a Polish space structure
(see e.g. \cite[Section 15.7.7]{Kal} and \cite[Proposition
3.17]{Res}), by means of which we introduce the Borel
$\sigma$-algebra $\mathcal{B}(\Gamma (X))$.  Then we fix a
probability measure $\mu$ on $\mathcal{B}(\Gamma (X))$ and interpret
the underlying set $\gamma $ as a realization of a random point
process. A typical choice of $\mu $ is a Poisson measure. However,
our results are valid for a wide class of probability measures on
$\Gamma (X)$ introduced below.
 Our goal is to study the set $\mathcal{G}(S^{\gamma })$ of all
Gibbs measures associated with the collections of  $W_{xy}$ and
$\chi_x$, for $\mu $-almost all configurations $\gamma $. In the
physical terminology, cf. \cite{Bov}, the elements of
$\mathcal{G}(S^{\gamma })$ are {\it quenched Gibbs states} of an
amorphous magnet. Here we discuss the questions of existence of such
states and their measurable dependence on $\gamma $, while the
complementary paper \cite{DKKP} is devoted to the problem of phase
transitions in a more specific (ferromagnetic) version of this
model.

We assume that the interaction potentials have finite range, that
is, they satisfy condition $W_{xy}\equiv 0$ whenever $\left\vert
x-y\right\vert >R$ for some fixed $R>0$. This allows for introducing
a graph structure on $\gamma $ in the following way: the vertex set
of the graph is $\gamma$ itself, whereas the edge set is defined as
\begin{equation}
\mathcal{E}_{\gamma }=\left\{ \{x,y\}\subset \gamma :\left\vert
x-y\right\vert \leq R\right\} .  \label{graph}
\end{equation}%
Correspondingly, vertices $x$ and $y$  are called neighbors or
adjacent in $\gamma $, if $|x-y|\leq R$. Let $n_{\gamma ,R}(x)$
denote the degree of vertex $x\in \gamma $, i.e., the number of
neighbors of $x$ in $\gamma $. In our model the function $n_{\gamma
,R}$ appears to be unbounded for $\mu $-a.a. $\gamma \in \Gamma
(X)$.  For unbounded degree graphs  and unbounded spins, the
question of existence of Gibbs measures was first studied in
\cite{KKP1}, where certain growth conditions on the degrees and
stability conditions on $\chi $ and $W_{xy}$ were imposed. Observe
that in the case of a compact spin space $S$ the answer to the
existence question is always positive, see e.g.  \cite[Proposition
5.3]{Pre}. For a comprehensive review of the theory of Gibbs
measures on graphs see \cite{GHM} and references therein.

The structure of this paper is as follows. In Section
\ref{estimates}, we describe certain properties of the graph defined
in (\ref{graph}), which hold for $\mu $-almost all $\gamma $. The
only assumption is that the measure $\mu $ has correlation functions
up to a certain order, which are
essentially bounded. In particular, we obtain bounds on the growth of $%
n_{\gamma ,R}(x)$ and show that, for $\mu $-a.a. $\gamma \in \Gamma
(X)$, the graph $(\gamma ,\mathcal{E}_{\gamma })$ satisfies the
corresponding conditions of work \cite{KKP1}. In Section
\ref{gibbs0}, by means of the results of \cite{KKP1} we show that
the set $\mathcal{G}(S^{\mathcal{\gamma }})$ is non-empty. In
addition, we describe the support of the elements of
$\mathcal{G}(S^{\mathcal{\gamma }})$ and obtain uniform estimates on
their exponential moments. By means of these estimates, we prove
that $\mathcal{G}(S^{\mathcal{\gamma }})$ contains elements with a
priori prescribed support properties. These are tempered Gibbs
measures. The use of such measures is typical for systems of
unbounded spins. The proof of these results is based on exponential
moment bounds for the local Gibbs specification of our model  and
its weak dependence on the boundary conditions. Such a technique is
effective in dealing with spatially irregular systems, see
\cite{KKP1,KKP2,KPR}. The two fundamental tools -- Ruelle's
(super-)\thinspace stability technique \cite{Ru69,Ru70} and general
Dobrushin's existence and uniqueness criteria \cite{Do70} -- are not
directly applicable to our model (due to the unboundedness of the
degree function $n_{\gamma ,R}$ and the lack of the spatial
transitivity of $\mu$-almost all $\gamma $). At the same time, for
our model the uniqueness problem remains open. Thus, the map $\Gamma
(X)$ $\ni \gamma \rightarrow \nu _{\gamma }\in \mathcal{G}(S^{\gamma
})$ is in general set-valued.

The results of Section \ref{gibbs0}, however, do not answer the
following important
question: is it possible to select $\nu _{\gamma }\in \mathcal{G}%
(S^{\gamma })$ in such a way that the resulting map $\gamma \mapsto
\nu _{\gamma }$ is measurable (existence of measurable selections)?
This
measurability is a key property that allows one to define averages of the type of $\int_{\Gamma (X)}\Phi (%
\mathbb{E}_{\nu _{\gamma }}F)\mu (d\gamma )$. Similar problems
appear in the theory of Gibbs fields with random components, e.g.,
random interactions. The measurable maps $\gamma \mapsto \nu
_{\gamma }$ are then called random Gibbs measures, see \cite{KKP2}
and Section 6.2 in \cite{Bov}. In Section \ref{marked}, we prove the
existence of random Gibbs measures for our model. For unbounded
spins with random interactions on a lattice, a similar result was
obtained in \cite{KKP2}. The novelty of the present
situation is that the measures $\nu _{\gamma }$ live (for different $\gamma $%
) on different spaces, and a priori it is not clear in what sense
the mentioned measurability can be understood. In Section
\ref{marked}, we develop a constructive
procedure of obtaining measurable selections $\gamma \mapsto \nu _{\gamma }$%
. For this, we identify the spaces $S^{\gamma }$, $\gamma \in \Gamma
(X)$, with the fibres of a
natural bundle over $\Gamma (X)$. It turns out that its total space $%
\mathfrak{X}$ has the structure of the marked configuration space
$\Gamma (X,S)$. For definitions and main facts on marked
configuration spaces we refer to \cite{DVJ1,AKLU,CG,KunaPhD}. Using
the appropriate moment bounds, we construct an auxiliary measure
$\hat{\nu}$ on $\Gamma (X,S)$ and define its conditional
distribution (i.e. disintegration) $\left( \nu _{\gamma }\right)
_{\gamma \in \Gamma (X)}\subset \mathcal{P}(S^{\gamma })$ with
respect to $\mu $, so that the measurability required holds. Then we
prove that $\nu _{\gamma }\ \in \mathcal{G}(S^{\gamma })$ and that
each $\nu _{\gamma }$ is a tempered measure. Note that $\nu _{\gamma
}$ need not in general coincide with the element of
$\mathcal{G}(S^{\gamma })$ constructed in the proof of Theorem
\ref{gibbs1} and be represented as the limit of a sequence of local
Gibbs measures. However, by means of Koml\'{o}s' theorem, we show
the existence (and hence measurability) of limiting Gibbs measures
$\nu _{\gamma }$ obtained from sequences of the Ces\`{a}ro means of
local Gibbs measures. It resembles the Newman--Stein approach
\cite{New,NS} in the theory of disordered spin systems, in which
the so called  `chaotic size dependence' is tamed by means of a
space averaging, see also \cite{KKP2}.

\section{Estimates for a typical configuration. \label{estimates}}

Let $C_{0}(X)$ denote the set of all continuous functions on
$f:X\rightarrow \mathbb{R}$ which have compact support. The
configuration space $\Gamma (X)$ defined in (\ref{gamma}) is
equipped with the vague topology -- the weakest topology that makes
continuous all the mappings
\begin{equation*}
\Gamma (X)\ni \gamma \mapsto \left\langle f,\gamma \right\rangle
:=\sum_{x\in \gamma }f(x),\quad f\in C_{0}(X).
\end{equation*}%
It is known that this topology is completely metrizable, which makes
$\Gamma
(X)$ a Polish space (see, e.g., \cite[Section 15.7.7]{Kal} or \cite[%
Proposition 3.17]{Res}); an explicit construction of the appropriate
metric can be found in \cite{KKut}. By $\mathcal{P}(\Gamma (X))$ we
denote the
space of all probability measures on the corresponding Borel $\sigma $-algebra $\mathcal{B}%
(\Gamma (X))$. We will also use the algebra $\mathcal{B}_{0}(\Gamma (X))$ of local sets, $\mathcal{B}%
_{0}(\Gamma (X)):=\cup _{\Lambda \in
\mathcal{B}_{0}(X)}\mathcal{B}(\Gamma
(\Lambda ))$. The space of $\mathcal{B}(\Gamma (X))$ (resp. $\mathcal{B}%
_{0}(\Gamma (X))$) measurable bounded functions $f:\Gamma
(X)\rightarrow
\mathbb{R}$ will be denoted by $L^{\infty }(\Gamma (X))$ (resp. $%
L_{0}^{\infty }(\Gamma (X))$).

For a given $\mu \in \mathcal{P}(\Gamma (X))$, a measurable
symmetric
(w.r.t. permutations of its arguments) function%
\begin{equation*}
0\leq k_{m}:X^{m}\rightarrow \mathbb{R},\ \text{\ \ }m\in
\mathbb{N},
\end{equation*}%
is called the $m$-th order \textit{correlation function} of $\mu $
if for any non-negative measurable symmetric function
$g:X^{m}\rightarrow \mathbb{R}
$ the following holds%
\begin{eqnarray}
&&\int_{\Gamma (X)}\sum_{\left\{ x_{1},...,x_{m}\right\} \subset
\gamma
}g(x_{1},...,x_{m})\mu (d\gamma )  \label{corr-funct} \\
&=&\frac{1}{m!}%
\int_{X^{m}}g(x_{1},...,x_{m})k_{m}(x_{1},...,x_{m})dx_{1}...dx_{m}.
\notag
\end{eqnarray}%
From now on we assume that $\mu $ is fixed and that it has all
correlation functions up to some order $M\in \mathbb{N}$, which are
essentially bounded, i.e.,
\begin{equation}
||k_{m}||_{\infty }:=\mathrm{ess~sup}_{X^{m}}\text{\thinspace }%
k_{m}(x_{1},...,x_{m})<\infty ,\text{ \ \ }1\leq m\leq M.
\label{corr-b}
\end{equation}

\begin{rem}
\label{rem1} In the theory of random point processes, correlation
functions $k_{m}$ appear as densities (w.r.t. $dx_{1}...dx_{m}$) of
the so-called $m$-th factorial moment measures corresponding to $\mu
$ (see e.g. \cite[Section 5.4]{DVJ1}). The boundedness as in
(\ref{corr-b}) holds for a wide class of measures on $\Gamma (X)$
and implies the finiteness of local moments, i.e.,
\begin{equation*}
\int\limits_{\Gamma (X)}\left\vert \left\langle f,\gamma
\right\rangle
\right\vert ^{m}\mu (d\gamma )<\infty ,\text{ \ \ \ }f\in C_{0}(X),\text{ \ }%
m\le M.
\end{equation*}
For the standard Poisson point process $\mu =\pi _{z}$ with the
activity parameter
$z>0$ and Lebesgue intensity measure $zdx$, the correlation functions $%
k_{m}(x_{1},...,x_{m})$ are just constants $z^{m},$ $m\in
\mathbb{N}$. If there exists $\zeta >0$ such that $||k_{m}||_{\infty
}\leq \zeta ^{m}$ for all $m\in \mathbb{N}$, we say that the
correlations functions $k_{m}$ are sub-Poissonian or satisfy
Ruelle's bound. Such measures typically arise in classical
statistical mechanics as Gibbs modifications of the Poisson measure
$\pi _{z}$ by means of stable interactions, see \cite{Ru69,Ru70}.
Note that any  $\mu \in \mathcal{P}(\Gamma (X))$ such that
$(k_{m})_{m\in \mathbb{N}}\leq \zeta^m$ for all $m\in \mathbb{N}$,
is uniquely determined by its correlation functions.
General criteria allowing for reconstructing a state $\mu \in \mathcal{P%
}(\Gamma (X))$ from a given system of functions $(k_{m})_{m\in
\mathbb{N}}$ were established in \cite{AKLU,KKun,Len73}.
\end{rem}

Now let us turn to the graph $(\gamma ,\mathcal{E}_{\gamma })$
defined in (\ref{graph}). For $x\in \gamma $, its degree in this
graph is
\begin{equation*}
n_{\gamma ,R}(x):=N\left( \left\{ y\in \gamma :~y\sim x\right\}
\right) \in \mathbb{Z}_{+}:=\mathbb{N}\cup \{0\},
\end{equation*}%
where $x\sim y$ means that $\{x,y\}\in \mathcal{E}_{\gamma }$. For
$\alpha ,r>0$, we introduce weights
\begin{equation*}
w_{\alpha }(x):=e^{-\alpha \left\vert x\right\vert },\qquad x\in X,
\end{equation*}%
and consider the following functions on $\Gamma (X)$:
\begin{equation*}
a_{\alpha ,r}(\gamma ) :=\sum_{\{x,y\}\in \mathcal{E}_{\gamma
}}w_{\alpha }(x)\left[ n_{\gamma ,R}(x)n_{\gamma ,R}(y)\right] ^{r},
\quad r\geq 0,
\end{equation*}
\begin{equation}\label{balpha}
b_{\alpha }(\gamma ) :=\sum_{x\in \gamma }w_{\alpha
}(x)=\left\langle w_{\alpha },\gamma \right\rangle .
\end{equation}
Standard arguments (based on $n$-particle expansions) show that
$a_{\alpha ,r}$ and $b_{\alpha }$ are $\mathcal{B}(\Gamma
(X))$-measurable.
\begin{prop}
\label{prop1} Let $\mu $ be such that (\ref{corr-b}) holds with some integer $%
M\geq 2$. Then, for any $\alpha >0$ and $0\leq r\leq M/2-1$, we have
inclusions $a_{\alpha ,r}$, $b_{\alpha }\in L^{1}(\Gamma (X),\mu )$.
\end{prop}

\noindent \textbf{Proof. }1) Applying (\ref{corr-funct}) to $w_{\alpha }\in L^{1}(X)$ we obtain%
\begin{eqnarray*}
\int_{\Gamma (X)}b_{\alpha }(\gamma )\ \mu (d\gamma )
&=&\int_{\Gamma
(X)}\sum_{x\in \gamma }w_{\alpha }(x)\ \mu (d\gamma ) \\
&=&\int_{X}w_{\alpha }(x)k_{1}(x)dx\leq ||k_{1}||_{\infty
}\int_{X}e^{-\alpha \left\vert x\right\vert }dx<\infty .
\end{eqnarray*}%
2) Since $n_{\gamma ,R}(x)n_{\gamma ,R}(y)$ is either $0$ or $\geq 1
$, we have $a_{\alpha ,r}(\gamma )\leq a_{\alpha ,r^{\prime
}}(\gamma )$
whenever $r\leq r^{\prime }$. Thus it is sufficient to prove the inclusion $%
a_{\alpha ,r}\in L^{1}(\Gamma (X),\mu )$ just for $r=M/2-1$ .

Let us fix some $x\in \gamma $. Clearly, for any $y\in \gamma $ such that $%
\left\vert x-y\right\vert \leq R$, we have%
\begin{equation*}
n_{\gamma ,R}(y)\leq n_{\gamma ,2R}(x),
\end{equation*}%
which yields
\begin{eqnarray}
\sum_{y\in \gamma \setminus \left\{ x\right\} }\left[ n_{\gamma
,R}(x)n_{\gamma ,R}(y)\right] ^{r} &\leq &n_{\gamma ,R}(x)\left[
n_{\gamma
,R}(x)n_{\gamma ,2R}(x)\right] ^{r}  \notag \\
&\leq &n_{\gamma ,2R}(x)^{2r+1}=n_{\gamma ,2R}(x)^{M-1}.
\label{ineq111}
\end{eqnarray}%
Observe that
\begin{equation*}
n_{\gamma ,2R}(x)=N\left( \left\{ y\in \gamma :0<\left\vert
x-y\right\vert
\leq 2R\right\} \right) =\sum_{\substack{ y\in \gamma  \\ y\neq x}}\mathbf{1}%
_{B_{2R}}(y-x),
\end{equation*}%
where $B_{2R}$ is the closed ball of radius $2R$ centred at the origin and $\mathbf{%
1}_{B_{2R}}$ is the corresponding indicator function. Thus, we have
the
multinomial expansion%
\begin{eqnarray*}
n_{\gamma ,2R}(x)^{M-1} &=&\left( \sum_{y\in \gamma \setminus
\left\{
x\right\} }\mathbf{1}_{B_{2R}}(y-x)\right) ^{M-1} \\
&=&\sum_{y_{1},...,y_{M-1}\in \gamma \setminus \left\{ x\right\}
}\prod_{k=1}^{M-1}\mathbf{1}_{B_{2R}}(y_{k}-x) \\
&=&\sum_{j=1}^{M-1}c_{j}\sum_{\left\{ y_{1},...,y_{j}\right\} \in
\gamma \setminus \left\{ x\right\}
}\prod_{k=1}^{j}\mathbf{1}_{B_{2R}}(y_{k}-x)
\end{eqnarray*}%
with the coefficients%
\begin{equation*}
c_{j}:=\sum\limits_{\substack{ i_{1},...,i_{j}\in \mathbb{N}  \\ %
i_{1}+...+i_{j}=M-1}}\frac{(M-1)!}{i_{1}!...i_{j}!},\text{ \ \
}1\leq j\leq M-1.
\end{equation*}%
Let us introduce notations $\bar{y}_{j}:=(y_{0},y_{1},...,y_{j})\in
\gamma ^{j+1}$ and $\left\{ \bar{y}_{j}\right\} :=\left\{
y_{0},y_{1},...,y_{j}\right\} \subset \gamma $ for the vector and
configuration with components $y_{0},$ $y_{1},...,y_{j}\in \gamma $,
respectively, and consider functions%
\begin{equation*}
f_{j}(\bar{y}_{j})=w_{\alpha }(y_{0})\prod_{k=1}^{j}\mathbf{1}%
_{B_{2R}}(y_{k}-y_{0})
\end{equation*}%
and%
\begin{equation*}
\hat{f}_{j}(\bar{y}_{j})=\sum_{s\in S_{j+1}}f_{j}(s(\bar{y}_{j})),
\end{equation*}%
where $S_{m}$ is the symmetric group of order $m$. Inequality (\ref{ineq111}) implies that%
\begin{equation}
a_{\alpha ,r}(\gamma )\leq \sum_{x\in \gamma }w_{\alpha
}(x)n_{\gamma ,2R}(x)^{M-1}=\sum_{j=1}^{M-1}c_{j}\sum_{\left\{
\bar{y}_{j}\right\} \subset \gamma }\hat{f}_{j}(\bar{y}_{j}).
\label{ineq112}
\end{equation}%
The application of (\ref{corr-funct}) to the right-hand side of (\ref{ineq112}) shows that%
\begin{equation*}
\int_{\Gamma (X)}a_{\alpha ,r}(\gamma )\mu (d\gamma )\leq \sum_{j=1}^{M-1}%
\frac{c_{j}}{(j+1)!}\int_{X^{j+1}}\hat{f}_{j}(\bar{y}_{j})k_{j+1}(\bar{y}%
_{j})d\bar{y}_{j}.
\end{equation*}%
Thus we obtain the estimate
\begin{multline*}
\int_{\Gamma (X)}a_{\alpha ,r}(\gamma )\mu (d\gamma )\leq \sum_{j=1}^{M-1}%
\frac{c_{j}\left\vert S_{j+1}\right\vert
}{(j+1)!}\int_{X^{j+1}}w_{\alpha
}(y_{0})\prod_{k=1}^{j}\mathbf{1}_{B_{2R}}(y_{k}-y_{0})k_{j+1}(\bar{y}_{j})d%
\bar{y}_{j} \\
\leq ||k||_{\infty }\sum_{j=1}^{M-1}c_{j}\mathrm{Vol}(B_{2R})^{j}%
\int_{X}e^{-\alpha \left\vert x\right\vert }dx<\infty ,
\end{multline*}%
where $\mathrm{Vol}(B_{2R})$ is the volume of the ball $B_{2R}$ and $%
||k||_{\infty }:=\underset{1\leq m\leq M}{\max }||k_{m}||_{\infty
}$, which completes the proof. $\square $

\section{Construction of Gibbs measures \label{gibbs0}}

In the standard \emph{%
Dobrushin-Lanford-Ruelle\ approach} in statistical mechanics
\cite{Geor,Pre} which we follow in this work, Gibbs states are
constructed by means of their local conditional distributions
(constituting the so-called Gibbsian specification). The main
technical problem in realizing this approach is to control the
spatial irregularity of the configuration $\gamma $ and the
unboundedness of the interaction potentials $W_{xy}$.

In what follows, we write $|\cdot |$ for the corresponding Euclidean
norms in both $X$ and $S$. Let $W_{xy}:S\times S\rightarrow
\mathbb{R}$, $x,y\in X$, be measurable functions satisfying the
polynomial growth estimate
\begin{equation}
\left\vert W_{xy}(u,v)\right\vert \leq I_{W}\left( \left\vert
u\right\vert ^{r}+\left\vert v\right\vert ^{r}\right) +J_{W},\ \ \
u,v\in S, \label{W-est}
\end{equation}%
and the finite range condition $W_{xy}\equiv 0$ if $\left\vert
x-y\right\vert \leq R$ for all $x,y\in X$ and some constants $%
I_{W},J_{W},R,r\geq 0$. We assume also that $W_{xy}(u,v)$ is
symmetric with respect to the permutation of $(x,u)$ and $(y,v)$. A
typical example is given by the bilinear form
\begin{equation}
W_{xy}(u,v)=A(x-y)u\cdot v,\ \text{\ }u,v\in S,  \label{W-quadr}
\end{equation}%
where $\cdot $ denotes the Euclidean inner product in $S$ and $A$ is
a uniformly bounded measurable mapping with values in the space of
symmetric $m\times m$ matrices such that $\mathrm{supp~}A\subset
B_{R}:=\left\{ x\in X\text{ }:\text{ }|x|\leq R\right\}$.

In the sequel, we take the single-spin measures in the following
form
\begin{equation*}
\chi _{x}(du):=e^{-V(u)}du,
\end{equation*}%
where $V:S\rightarrow \mathbb{R}$ is a measurable functions
satisfying
\begin{equation}
V(u)\geq a_{V}\left\vert u\right\vert ^{q}-b_{V},\ \ u\in S,
\label{V-est}
\end{equation}%
for some constants $a_{V},b_{V}>0$ , and $q>2$ Note that $\chi
_{x}(S)<\infty $ in view of (\ref{V-est}), which is aimed to
compensate the destabilizing effects
of the unbounded interactions potential $W_{xy}$. Note also that the case of $%
q=2 $ cannot be covered by our scheme due to the lack of uniform
bounds on vertex degrees $n_{\gamma ,R}(x)$ in the underlying graph $(%
\mathcal{\gamma },\mathcal{E}_{\gamma })$\textbf{. }

For a fixed $\gamma \in \Gamma (X)$, we will denote by $\sigma
_{\gamma },\xi _{\gamma }$, etc. elements of the space $S^{\gamma
}$, and omit the subscript $\gamma $ whenever possible. Let
$\mathcal{F}(\gamma )$ be the
collection of all finite subsets of $\gamma $. For any $\eta \in \mathcal{F}%
(\gamma ),$ $\sigma _{\eta }=(\sigma (x))_{x\in \eta }\in S^{\eta }$ and $%
\xi _{\gamma }=(\xi (y))_{y\in \gamma }\in S^{\gamma }$ define the
relative
local interaction energy%
\begin{equation*}
E_{\eta }(\sigma \left\vert \xi \right. )=\sum_{\{x,y\}\subset \eta
}W_{xy}(\sigma (x),\sigma (y))+\sum_{\substack{ x\in \eta  \\ y\in
\gamma
\setminus \eta }}W_{xy}(\sigma (x),\xi (y)). 
\end{equation*}%
The corresponding specification kernel $\Pi _{\eta }(d\sigma
\left\vert \xi
\right. )\in \mathcal{P}(S^{\gamma })$ is given by the formula%
\begin{equation}
\int_{S^{\gamma }}f(\sigma )\Pi _{\eta }(d\sigma |\xi )=Z(\xi
)^{-1}\int_{S^{\eta }}f(\sigma _{\eta }\times \xi _{\gamma \setminus \eta })%
\mathrm{exp}~\left[ -E_{\eta }(\sigma \left\vert \xi \right.
)\right] \chi _{\eta }(d\sigma _{\eta }),  \label{spec-kernel}
\end{equation}%
where
\begin{equation}
\chi _{\eta }(d\sigma _{\eta }):=\bigotimes_{x\in \eta }\chi
_{x}(d\sigma (x)),  \label{measure-chi}
\end{equation}%
$f\in L^{\infty }(S^{\gamma })$ ( $=:$ the set of bounded Borel function on $%
S^{\gamma }$) and
\begin{equation*}
Z(\xi )=\int_{S^{\eta }}\mathrm{exp}~\left[ -E_{\eta }(\sigma
\left\vert \xi \right. )\right] \chi _{\eta }(d\sigma _{\eta })
\end{equation*}%
is a normalizing factor. Observe that the integral in the right-hand
side of (\ref{spec-kernel}) is well-defined in view of
(\ref{V-est}). For each fixed $\xi \in S^{\gamma }$, $\Pi _{\eta
}(d\sigma |\xi )$ is a
probability measure on $S^{\gamma }$ and, for each fixed $B\in \mathcal{B}%
(S^{\gamma })$, the map $S^{\gamma }\ni \xi \rightarrow \Pi _{\eta
}(B|\xi )\in \lbrack 0,1]$ is measurable. The family $\Pi _{\gamma
}:=\left\{ \Pi _{\eta }(d\sigma |\xi )\right\} _{\eta \in
\mathcal{F}(\gamma ),\xi \in S^{\gamma }}$ is called the Gibbsian
specification (see e.g. \cite{Geor,Pre}).
By construction, it satisfies the consistency property%
\begin{equation}
\int_{S^{\gamma }}\Pi _{\eta _{1}}(B\left\vert \sigma \right. )\Pi
_{\eta _{2}}(d\sigma \left\vert \xi \right. )=\Pi _{\eta
_{2}}(B\left\vert \xi \right. ),  \label{specification}
\end{equation}%
which holds for any $B\in \mathcal{B}(S^{\gamma })$, $\xi \in
S^{\gamma }$ and $\eta _{1},\eta _{2}\in \mathcal{F}(\gamma )$ such
that $\eta _{1}\subset \eta _{2}$.

A probability measure $\nu $ on $S^{\gamma }$ is said to be a  Gibbs
measure associated with the
potentials $W$ and $V$ if it satisfies the DLR equation%
\begin{equation}
\nu (B)=\int_{S^{\gamma }}\Pi _{\eta }(B|\xi )\nu (d\xi )
\label{DLR}, \quad B\in \mathcal{B}(S^{\gamma }),
\end{equation}%
for all $\eta \in \mathcal{F}(\gamma )$. Equivalently, one can fix
an exhausting sequence $(\Lambda_N)$ of compact sets in $X$ and
require (\ref{DLR}) only for
$\eta=\gamma\cap\Lambda_N,\,N\in{\mathbb{N}}$. For a given $\gamma
\in \Gamma (X)$, by $\mathcal{G}(S^{\gamma })$ we denote the set of
all such measures.

Our next goal is to prove the existence of Gibbs measures supported
on
certain sets of tempered\textit{\ }sequences from $S^{\gamma }$ for $\mu $%
-a.a. $\gamma \in \Gamma (X)$. Let us assume that the measure $\mu $
satisfies (\ref{corr-b}) with an integer $M$ (cf. Proposition \ref%
{prop1}) such that
\begin{equation}
M>\frac{2q}{q-2}>2,  \label{corr-c}
\end{equation}%
where $q$ is the same as in (\ref{V-est}). Fix a parameter
\begin{equation}
p\in \left[ \frac{2M}{M-2}\text{ },\text{ }q\right]
\label{parameters}
\end{equation}%
and set%
\begin{equation*}
p^{\prime }:=2\left( p-2\right) ^{-1},
\end{equation*}%
so that
\begin{equation*}
\frac{2}{q-2}\leq p^{\prime }\leq M/2-1.
\end{equation*}%
Thus, according to Proposition \ref{prop1}, $a_{\alpha ,p^{\prime
}},b_{\alpha }\in L^{1}(\Gamma (X),\mu )$ for any $\alpha >0$, and
thus
\begin{equation}
a_{\alpha ,p^{\prime }}(\gamma ),\ b_{\alpha }(\gamma )<\infty
\label{ab-ineq}
\end{equation}%
for $\mu $-a.a. $\gamma \in \Gamma (X)$.

For $\sigma \in S^{\gamma }$, we define the norm
\begin{equation}
\left\Vert \sigma \right\Vert _{\alpha ,p}:=\left( \sum_{x\in \gamma
}\left\vert \sigma (x)\right\vert ^{p}w_{\alpha }(x)\right) ^{1/p}
\label{norm-sigma}
\end{equation}%
and consider the Banach space%
\begin{equation*}
l_{\alpha }^{p}(\gamma ,S):=\left\{ \sigma \in S^{\gamma
}:\left\Vert \sigma \right\Vert _{\alpha ,p}<\infty \right\} .
\end{equation*}%
By $\mathcal{G}_{\alpha ,p}(S^{\gamma })\subset \mathcal{G}%
(S^{\gamma })$ we denote the set of all Gibbs measures on $\gamma $
associated with $W$ and $V$, which are supported on $l_{\alpha
}^{p}(\gamma ,S)$. These measures are called \textit{tempered}.

\begin{theorem}
\label{gibbs1}Assume that conditions (\ref{corr-c}) and
(\ref{parameters}) are satisfied. Then the following statements hold
for $\mu $-a.a. $\gamma \in \Gamma (X)$:\newline \ \ 1) the set
$\mathcal{G}_{\alpha ,p}(S^{\gamma })$ is not empty;\newline \ \ 2)
for any $\lambda \in \mathbb{R}_{+}$, there exists a constant $\Xi
_{\gamma }(\lambda )>0$ such that every $\nu \in \mathcal{G}_{\alpha
,p}(S^{\gamma })$ satisfies the estimate%
\begin{equation}
\int_{S^{\gamma }}\mathrm{exp}~\left\{ \lambda \left\Vert \sigma
\right\Vert _{\alpha ,p}^{p}\right\} ~\nu (d\sigma )\leq
\mathrm{exp}~\Xi _{\gamma }(\lambda ).  \label{exp-moment1}
\end{equation}
\end{theorem}

\noindent \textbf{Proof.} For $\gamma \in \Gamma (X)$ satisfying (\ref%
{ab-ineq}), statements 1) and 2) follow by the direct application of
Theorem 1 of \cite{KKP1} to the graph $(\gamma ,\mathcal{E}_{\gamma
})$. The key technical step is to establish the exponential moment
bound
\begin{equation*}
\mathrm{sup}_{\eta \in \mathcal{F}(\gamma )}\int_{S^{\gamma
}}e^{\lambda \left\Vert \sigma \right\Vert _{\alpha ,p}^{p}}\Pi
_{\eta }(d\sigma
\left\vert \xi \right. )< \infty ,\,\xi \in l_{\alpha }^{p}(\gamma ,S),  
\end{equation*}%
which holds uniformly in $\eta \in \mathcal{F}(\gamma )$ and implies
the \textit{local equicontinuity} of the family \newline $\left\{
\Pi
_{\eta }(d\sigma \left\vert \xi \right. )\right\} _{\eta \in \mathcal{F}%
(\gamma )}$ for any  $\xi \in l_{\alpha }^{p}(\gamma ,S)$ (cf.
Definition 4.6 in
\cite{Geor}) and hence its relative compactness in the topology of \textit{%
set-wise convergence}
on the algebra $\mathcal{B}_{0}(S^{\gamma }):=\cup _{\eta \in \mathcal{F}%
(\gamma )}\mathcal{B}_{\eta }(S^{\gamma })$ of local subsets of $S^{\gamma }$%
. Here $\mathcal{B}_{\eta }(S^{\gamma })$ is the $\sigma $-algebra
of sets $C_A=
\left\{ \sigma \in \mathcal{B}(S^{\gamma }):\sigma _{\eta }\in A\right\} $, $A\in\mathcal{%
B}(S^{\eta })$, which is isomorphic to $\mathcal{B}(S^{\eta })$.

 This ensures the existence of accumulation points $\nu ^{\xi
}\in \mathcal{P}(S^{\gamma })$. Standard limit transition arguments
show that $\nu ^{\xi }\in \mathcal{G}_{\alpha ,p}(S^{\gamma })$, and
that estimate (\ref{exp-moment1}) holds for all $\nu \in
\mathcal{G}_{\alpha ,p}(S^{\gamma })$.

The only additional features of the present framework (in comaprison to \cite%
{KKP1}) are the multidimensionalinty of the spin space $S$ and the
dependence of the potentials $W_{xy}$ on $x,y\in \gamma $, which
however does not affect the proof in view of the uniformity of the
estimates in (\ref{W-est}). $\square $

The next important auxiliary statement is a byproduct of the proof
of Lemma 1 in \cite{KKP1}.
\begin{prop}
\label{moments copy(1)}For each $\lambda >0$, $\beta \in
(0,e^{\alpha R}\lambda /2)$ and\textbf{\ }$p^{\prime }=2(p-2)^{-1}$,
there exist constants
$C_{1},C_{2},C_{3}\geq 0$ such that the following estimate holds:%
\begin{equation}
\int_{S^{\gamma }}\left\Vert \sigma \right\Vert _{\alpha ,p}^{p}\Pi
_{\eta }(d\sigma \left\vert \xi \right. )\leq C_{1}b_{\alpha
}(\gamma )+C_{2}a_{\alpha ,p^{\prime }}\left( \gamma \right)
+C_{3}\left\Vert \xi _{\gamma \setminus \eta }\right\Vert _{\alpha
,p}^{p},  \label{moment0}
\end{equation}%
uniformly for all $\eta \in \mathcal{F}(\gamma )$. Moreover,
\begin{equation}
\int_{S^{\gamma }}\left\Vert \sigma \right\Vert _{\alpha ,p}^{p}~\nu
(d\sigma )\leq C_{1}b_{\alpha }(\gamma )+C_{2}a_{\alpha ,p^{\prime
}}\left( \gamma \right) , \label{moment-est}
\end{equation}%
holding for any $\nu \in \mathcal{G}_{\alpha ,p}(S^{\gamma })$.
\end{prop}

\noindent \textbf{Proof. }The application of Jensen's inequality to
the right-hand side of formula (3.6) of \cite{KKP1} together with
(3.18) of the same work implies (\ref{moment0}). Bound
(\ref{moment-est}) can be proved by limit transition arguments
combined with the DLR\ equation, similar to the proof of
(\ref{exp-moment1}).
\hfill%
$\square $

\begin{rem}
Condition (\ref{corr-c}) establishes a relation between the growth
rate $q$ of  $V$ and the number $M$ related to the correlation
functions $k_{m},\ 1\leq m\leq M$,\ of the underlying random point
process $\mu $. In the case where $\mu $ has bounded correlation
functions of arbitrary order, condition (\ref{corr-c}) holds for any
$q>2$ and $p\in \left( 2,q\right] $. Observe that higher values of
$p$ guarantee the existence of Gibbs measures with the smaller
support set $l_{\alpha }^{p}(\gamma ,S)$.\textbf{\ }Unfortunately,
our method does not allow us to
control the case of $q=2,$ even when the underlying particle configuration $%
\gamma $ is distributed according to the homogeneous Poisson random field $%
\mu =\pi _{z}$ on $\Gamma (X)$. In particular, the existence problem
is
still open\ for the important class of ferromagnetic harmonic systems on $%
S^{\gamma }$ with the pair interactions of the form (\ref%
{W-quadr}) with $A(x-y)\leq 0$ and $V(u)=a_{V}\left\vert
u\right\vert ^{2},$ $a_{V}>0.$
\end{rem}

\begin{rem}
As already mentioned in the Introduction, in this paper we do not
touch the question of uniqueness of $\nu \in \mathcal{G}(S^{\gamma
})$. This is a
highly non-trivial problem and general conditions that guarantee that $N(%
\mathcal{G}(S^{\gamma }))=1$ are not known (even for small
interaction strength). On the other hand, in \cite{DKKP} we studied
a class of models with ferromagnetic pair interaction living on
Poisson random graphs and
showed the existence of multiple Gibbs states, that is, that $N(\mathcal{G}%
(S^{\gamma }))>1$ (and therefore $=\infty $) for a.a. $\gamma \in \Gamma (X)$%
.
\end{rem}

\section{Measurable dependence on $\protect\gamma $. \label{marked}}

In the proof of Theorem \ref%
{gibbs1}, a measure $\nu _{\gamma }^{\xi }\in \mathcal{G}_{\alpha
,p}(S^{\gamma })$ has been constructed for each tempered $\xi \in
\Gamma (X)$ as a limit of a sequence of 'finite volume'\ measures
$\Pi _{\eta _{n}}(d\sigma |\xi )$, $n\in \mathbb{N}$. However, the
measurability of the map $\Gamma (X)\ni \gamma \mapsto \nu _{\gamma
}^{\xi }$ is far from being clear. Indeed, the sequence
$\mathbf{\eta }=\left( \eta _{n}\right) _{n\in \mathbb{N}}\subset
\mathcal{F}(\gamma )$ can depend on the random parameter\ $\gamma $
in some uncontrollable way (the so-called
\emph{chaotic} \emph{size} \emph{dependence}, see the discussion in \cite%
{New,NS}).  In this section  we address this problem. A difficulty
here is
 that, for different $\gamma $, the measures $\nu _{\gamma }\in
\mathcal{G}_{\alpha ,p}(S^{\gamma })$ are defined on different
spaces, and it is not clear in what sense this measurability can be
understood. To overcome this difficulty we will identify the spaces
$S^{\gamma }$ that support measures $\nu _{\gamma }\in
\mathcal{G}_{\alpha ,p}(S^{\gamma })$ with the measurable subspaces
of the \textit{marked configuration space}
\begin{equation}
\Gamma (X,S):=\left\{ \hat{\gamma}\in \Gamma (X\times S):p_{X}(\hat{\gamma}%
)\in \Gamma (X)\right\} ,  \label{def-marked}
\end{equation}%
where $p_{X}$ is the natural extension to $\Gamma (X\times S)$ of
the canonical projection $X\times S\rightarrow S$. For basic
definitions and properties of marked configuration spaces we refer
to e.g. \cite{AKLU,CG,DVJ1,KunaPhD}.

In order to proceed, we  endow the space $\Gamma (X,S)$ with
a\textbf{\ (}completely metrizable\textbf{)} topology defined as the
weakest topology that makes the map   $\Gamma (X,S)\ni
\hat{\gamma}\mapsto \left\langle f,\hat{\gamma}\right\rangle$
continuous for any $f\in C_{\mathrm{b,}0}(X\times S)$ ( $=:$ the set
of continuous bounded functions on $X\times S$ with support
$S_{\Lambda
}:=\Lambda \times S$, $\Lambda \in \mathcal{B}_{0}(X)$). Let  $\mathcal{B}(\Gamma (X,S))$ be  the corresponding Borel $%
\sigma $-algebra. The space of $\mathcal{B}(\Gamma (X,S))$
measurable bounded functions $f:\Gamma (X,S)\rightarrow \mathbb{R}$
will be denoted by $L^{\infty }(\Gamma (X,S))$.

The space $\Gamma (X,S)$ has the structure of a fibre bundle over
$\Gamma (X) $, with fibres $p_{X}^{-1}(\gamma )$, which can be
identified with the product spaces $S^{\gamma }$. As before,
elements of $S^{\gamma }$ will be denoted by $\sigma \equiv \sigma
_{\gamma }=(\sigma (x))_{x\in \gamma }$,
with the subscript $\gamma $ omitted when possible. Thus each $\hat{\gamma}%
\in \Gamma (X,S)$ can be represented by the pair
\begin{equation*}
\hat{\gamma}=(\gamma ,~\sigma ),\text{ where }\gamma
=p_{X}(\hat{\gamma})\in \Gamma (X),~\sigma \in S^{\gamma }.
\end{equation*}%
It follows directly from the definition of the corresponding
topologies that the map $p_{X}:\Gamma (X,S)\rightarrow \Gamma (X)$
is continuous, which implies that the space $S^{\gamma }$ is a Borel
subset of $\Gamma (X,S)$ for any configuration $\gamma \in \Gamma
(X)$. Moreover,
  $S^\gamma$ is a Polish space embedded into $\Gamma (X,S)$, which is a Polish space as well.
By the Kuratowski theorem \cite[page 21]{Par}, the  latter implies
that the Borel $\sigma$-algebras  $\mathcal{B}(S^\gamma)$ and
\begin{equation*}
\mathcal{A}( \mathbb{R}^\gamma) := \{ A \in  \mathcal{B}(\Gamma
(X,S)): A \subset S^\gamma\}
\end{equation*}
are measurably isomorphic. Thus, any probability measure $\mu$ on
$\mathcal{B}(\Gamma (X,S))$ with the property $\mu(S^\gamma)=1$ can
be redefined as a measure on $\mathcal{B}(S^\gamma)$, for which we
will you the same notation.

Let $\mathcal{P}(\Gamma (X,S))$ stand for the space of all Borel
probability
measures on $\Gamma (X,S)$. We say that a map $\Gamma (X)\ni \gamma \mapsto \nu _{\gamma }\in \mathcal{P}%
(\Gamma (X,S))$ is measurable if  the map $\Gamma (X)\ni \gamma
\mapsto \nu _{\gamma }(A)\in\mathbb{R}$ is measurable for all $A\in
\mathcal{B}(\Gamma (X,S))$, which in turn is equivalent to the
measurability of the map $\Gamma (X)\ni \gamma \mapsto \int_{\Gamma
(X,S)}f(\hat{\eta})\nu _{\gamma }(d\hat{\eta})\in\mathbb{R}$ for all
$f\in L^\infty(\Gamma (X,S))$.

The following theorem is the main result of this section.

\begin{theorem}
\label{th-meas} There exists a measurable mapping%
\begin{equation}
\Gamma (X)\ni \gamma \mapsto \nu _{\gamma }\in \mathcal{P}(\Gamma
(X,S)) \label{meas}
\end{equation}%
such that $\nu _{\gamma }(S^\gamma)=1$ and
$\nu _{\gamma }\in \mathcal{G}_{\alpha ,p}(S^{\gamma })$ for $\mu $%
-a.a. $\gamma \in \Gamma (X)$.
\end{theorem}

The proof will go along the following lines. First, using moment bounds (\ref%
{moment0}), we will construct an auxiliary measure $\hat{\nu}$ on
$\Gamma (X,S)$ and define its conditional distribution (i.e.
disintegration) $\left( \nu _{\gamma }\right) _{\gamma \in \Gamma
(X)}\subset \mathcal{P}(S^{\gamma
})$ with respect to $\mu $, so that the measurability required in (\ref{meas}%
) holds. Then we will prove the inclusion $\nu _{\gamma }\in \mathcal{G}%
_{\alpha ,p}(S^{\gamma })$.

Let us fix a measurable mapping $u:X\rightarrow S$ satisfying the bound  $%
\left\vert u(x)\right\vert \leq ce^{\beta \left\vert x\right\vert }$
for some $c,\beta \in \mathbb{R}$ and define the map
\begin{equation}
\Gamma (X)\ni \gamma \mapsto \xi _{\gamma }=(u(x))_{x\in \gamma }\in
S^{\gamma }.  \label{section}
\end{equation}%
Obviously, we have the esimate
\begin{equation}
\left\Vert \xi _{\gamma }\right\Vert _{\alpha ,p}^{p}\leq b_{\alpha
^{\prime }}(\gamma ),\ \text{for any }\alpha ^{\prime }>p\beta ,
\label{sec-bound}
\end{equation}%
so that $\xi _{\gamma }\in l_{\alpha ^{\prime }}^{p}(\gamma ,S)$ for $\mu $%
-a.a. $\gamma \in \Gamma (X)$.

\begin{prop}
Let $\xi _{\gamma }$ be as in (\ref{section}). Then the map
\begin{equation*}
\Gamma (X)\ni \gamma \mapsto (\gamma ,\xi _{\gamma })\in \Gamma
(X,S)
\end{equation*}%
is measurable.
\end{prop}

\noindent \textbf{Proof. } By the\textbf{\ }definition of the
measurable structure of ~$\Gamma (X,S)$,\textbf{\ }the claim is
equivalent to the measurability of the maps $
\Gamma (x)\ni \gamma \mapsto F(\gamma ):=\left\langle f,\hat{\gamma}%
\right\rangle ,\ \ \hat{\gamma}=(\gamma ,\xi _{\gamma }), $ for all
$f\in C_{\mathrm{b},0}(X\times S)$. It is clear that $F(\gamma
)=\left\langle g,\gamma \right\rangle $, where $g(x)=f(x,u(x))$, so
that $g$ is measurable and has compact support. The assertion
follows now from the definition of the measurable structure of
$\Gamma (X)$.
\hfill%
$\square $\textbf{\ }

Let us fix $\Delta \in \mathcal{B}_{0}(X)$ (e.g. a
closed ball or cube) and define $\gamma _{\Delta }:=\Delta \cap \gamma $, $%
\gamma \in \Gamma (X)$. Obviously, $\gamma _{\Delta }\in \mathcal{F}(\gamma )$%
. Consider the measure $\hat{\nu}_{\Delta }^{\xi }$ on $\Gamma
(X,S)$ defined by the formula
\begin{equation}
\hat{\nu}_{\Delta }^{\xi }(d\hat{\gamma})=\Pi _{\gamma _{\Delta
}}(d\sigma
\left\vert \xi _{\gamma }\right. )\mu (d\gamma ),\ \ \gamma =p_{X}(\hat{%
\gamma}),  \label{nu-measure}
\end{equation}%
or, equivalently,
\begin{equation*}
\int_{\Gamma (X,S)}F(\hat{\gamma})\hat{\nu}_{\Delta }^{\xi }(d\hat{\gamma}%
)=\int_{\Gamma (X)}\Phi _{F}(\gamma )\mu (d\gamma )
\end{equation*}%
holding for each $F\in C_{b}(\Gamma (X,S))$, where%
\begin{equation*}
\Phi _{F}(\gamma ):=\int_{S^{\gamma }}F(\gamma ,\sigma )~\Pi
_{\gamma _{\Delta }}(d\sigma \left\vert \xi _{\gamma }\right. ).
\end{equation*}%
The measure $\hat{\nu}_{\Delta }^{\xi }$ is well-defined because of
the next result.

\begin{prop}
The function $\Phi _{F}:\Gamma (X)\rightarrow \mathbb{R}$ is
measurable.
\end{prop}

\noindent \textbf{Proof. } For any $A,B\subset X$, introduce the
function $W_{A,B}(x\times u,y\times v)=\mathbf{1}_{A
}(x)\mathbf{1}_{B }(y)W_{xy}(u,v)$, $x,y\in X$,$u,v\in S$, and set $
\widehat{W}_{\Delta }(\hat{\gamma}):=\sum_{\left\{ \hat{x},\hat{y}\right\} \subset \hat{\gamma}%
}\left( W_{\Delta \times \Delta }+W_{\Delta \times \Delta ^{c}}\right) (\hat{%
x},\hat{y}). $ Observe that $\widehat{W}_{\Delta }:\Gamma
(X,S)\rightarrow \mathbb{R}$ is measurable. For
$\hat{\gamma}=(\gamma ,\sigma _{\gamma _{\Delta }}\times \xi
_{\gamma _{\Delta ^{c}}})$ we have  the equality $
E_{\gamma _{\Delta }}(\sigma \left\vert \xi _{\gamma }\right. )=\widehat{W}%
_{\Delta }(\hat{\gamma}), $ which implies the measurability of the
map $\gamma\mapsto E_{\gamma _{\Delta }}(\sigma \left\vert \xi
_{\gamma }\right )$. It follows from (\ref{spec-kernel}) and
(\ref{measure-chi}) that
\begin{equation*}
\Phi _{F}(\gamma ) =Z^{-1}\int F(\hat{\gamma})\exp \left(-E_{\gamma
_{\Delta }}(\sigma \left\vert \xi _{\gamma }\right)\right) \chi
_{\gamma _{\Delta }}(d\sigma _{\gamma_{\Delta }}),
\end{equation*}
where $Z =\int \exp \left( -E_{\gamma _{\Delta }}(\sigma \left\vert
\xi _{\gamma }\right )\right) \chi _{\gamma _{\Delta }}(d\sigma
_{\gamma _{\Delta }}). $
 Without loss of generality we can assume that $\chi _{x}$ is a probability measure. It was proved in \cite[(2.18) and
Appendix A]{BoDal} that the map%
\begin{equation*}
\Gamma (X)\ni \gamma \mapsto \int G(\gamma ,\sigma )\chi _{\gamma
}(d\sigma )\in \mathbb{R}
\end{equation*}%
is measurable for any measurable function $G:\Gamma (X,S)\rightarrow \mathbb{%
R}$. The result follows now from the measurability of the map
 $\Gamma (X)\ni \gamma \mapsto
\gamma _{\Delta }\in \Gamma (\Delta )$.
\hfill%
$\square $

Let us consider the algebra $\mathcal{B}_{0}(\Gamma (X,S)):=\cup
_{\Lambda \in \mathcal{B}_{0}(X)}\mathcal{B}_{\Lambda }(\Gamma
(X,S))$ of local subsets of $\Gamma (X,S)$. Here
$\mathcal{B}_{\Lambda }(\Gamma (X,S))$ is the $\sigma $-algebra of
sets $C_{A}:=\left\{ \hat{\gamma}\in \Gamma (X,S):\hat{\gamma}\cap
S_{\Lambda }\in A\right\} $, $A\in \mathcal{B}(\Gamma (\Lambda
,S))$, which is isomorphic to $\mathcal{B}(\Gamma (\Lambda ,S))$.
The space of  $\mathcal{B}%
_{0}(\Gamma (X,S))$ measurable bounded functions $f:\Gamma
(X,S)\rightarrow
\mathbb{R}$ will be denoted by  $%
L_{0}^{\infty }(\Gamma (X,S))$.

Our next goal is to show that the family $\left\{ \hat{\nu}_{\Delta
}^{\xi },\ \Delta \in \mathcal{B}_{0}(X)\right\} \subset
\mathcal{P}(\Gamma (X,S))$ has an accumulation point. For this, we
equip the space $\mathcal{P}(\Gamma (X,S))$ with the topolgy $\tau
_{loc}$ of local setwise convergence (cf. \cite[Sec. 4.1, Prop.
4.9]{Geor}). This is the weakest topolgy that makes the maps
$\mathcal{P}(\Gamma (X,S))\ni \mu \mapsto \mu (B)\in \mathbb{R}$
(resp. $\mathcal{P}(\Gamma (X,S))\ni \mu \mapsto \int f(\hat{\gamma})\mu (d%
\hat{\gamma})\in \mathbb{R}$) continuous for all $B\in \mathcal{B}%
_{0}(\Gamma (X,S))$ (resp. all $\mathcal{B}_{0}(\Gamma
(X,S))$-measurable
bounded functions $f:\Gamma (X,S)\rightarrow \mathbb{R}$), so that $\mathcal{%
P}(\Gamma (X,S))\ni \mu _{n}\overset{\tau _{loc}}{\rightarrow }\mu
\in \mathcal{P}(\Gamma (X,S)),\ n\rightarrow \infty ,\ $iff $\mu
_{n}(B)\rightarrow \mu (B)\ $for all $B\in \mathcal{B}_{0}(\Gamma
(X,S))$.
\begin{definition}
\label{equi} We say that a family of probability measures $\left\{
\mu _{m}\right\} _{m\in \mathbb{N}}$ on $\Gamma (X,S)$ is locally
equicontinuous (LEC) if for any $\Delta \in \mathcal{B}_{0}(X)$ and
any sequence $\left\{
B_{n}\right\} _{n\in \mathbb{N}}\in \mathcal{B}(\Gamma (\Delta ,S))$, $%
B_{n}\downarrow \emptyset $, $n\rightarrow \infty $, we have%
\begin{equation}
\underset{n\rightarrow \infty }{\mathrm{lim}}\mathrm{\
}\underset{m\in \mathbb{N}}{\mathrm{lim~sup}}\ \mu _{m}\left(
B_{n}\right) =0. \label{equicont0}
\end{equation}
\end{definition}
Observe that the local setwise convergence is equivalent to
convergence in
the space $[0,1]^{\mathcal{B}_{0}}$, where $\mathcal{B}_{0}:=\mathcal{B}%
_{0}(\Gamma (X,S))$. The following fact is essentially well-known,
see \cite[Prop. 4.9]{Geor}.
\begin{prop}
\label{equicont1} Let $\left\{ \mu
_{n}\right\} _{n\in \mathbb{N}}$ be a LEC family of probability measures on $%
\Gamma (X,S).$ Then it has a $\tau _{loc}$-cluster point, which is
also a probability measure on $\Gamma (X,S)$.
\end{prop}

\noindent \textbf{Proof. } We give here a sketch of the proof from
\cite[Prop. 4.9]{Geor} adapted to our setting. It is straightforward
that the family $\left\{ \mu _{n}\right\} _{n\in \mathbb{N}}$
contains a cluster point $\mu $ as an element of the compact space
$[0,1]^{\mathcal{B}_{0}}$, and $\mu $ is an additive function on
$\mathcal{B}_{0}$. The LEC property (\ref{equicont0})
implies that the projection $\mu _{\Lambda }$ of $\mu $ onto $\mathcal{B}%
(\Gamma (\Lambda ,S))$ is $\sigma $-additive for each $\Lambda \in \mathcal{B%
}_{0}(X)$. Thus $\left\{ \mu _{\Lambda }\right\} _{\Lambda \in \mathcal{B}%
_{0}(X)}$ forms a consistent (w.r.t. projective maps $p_{\Lambda
_{2},\Lambda _{1}}:\Gamma (\Lambda _{2},S)\ni \hat{\gamma}\mapsto \hat{\gamma%
}_{\Lambda _{1}}:=(\gamma _{\Lambda _{1}},~\sigma _{\gamma _{\Lambda
_{1}}})\in \Gamma (\Lambda _{1},S)$, $\Lambda _{1}\subset \Lambda
_{2}$) family of measures and by the corresponding version of the
Kolmogorov theorem (see \cite[Theorem V.3.2 ]{Par})) generates a
probability measure on $\Gamma (X,S)$ (which obviously coincides
with $\mu $).
\hfill%
$\square $

\begin{corollary}
There exists a subsequence $\left\{ \mu _{n_{k}}\right\} _{k\in
\mathbb{N}}$
such that $\mu _{n_{k}}\overset{\tau _{loc}}{\rightarrow }\mu $, $%
k\rightarrow \infty $.
\end{corollary}

Let us consider the function
\begin{equation*}
\phi (\hat{\gamma}):=b_{\alpha}(\gamma) +\left\Vert \sigma _{\gamma
}\right\Vert _{\alpha ,p}^{p},\end{equation*}
 with $b_{\alpha}(\gamma)$ and $\left\Vert \sigma \right\Vert _{\alpha ,p}$ given
by formulae (\ref{balpha}) and  (\ref{norm-sigma}), respectively.
Using estimate (\ref{moment0}) we obtain the inequality
\begin{eqnarray*}
\int_{\Gamma (X,S)}\phi (\hat{\gamma})\hat{\nu}_{\Delta }^{\xi }(d\hat{\gamma%
}) &\leq &\int_{\Gamma (X)}\left[ (C_{1}+1)b_{\alpha }(\gamma
)+C_{2}a_{\alpha ,p^{\prime }}\left( \gamma \right) +C_{3}\left\Vert
\xi _{\gamma _{\Delta ^{c}}}\right\Vert _{\alpha ,p}^{p}\right] \mu
(d\gamma ).
\end{eqnarray*}%
It follows now from estimate (\ref{sec-bound}) and Proposition
\ref{prop1}  that
\begin{equation}
\int_{\Gamma (X,S)}\phi (\hat{\gamma})\hat{\nu}_{\Delta }^{\xi }(d\hat{\gamma%
})\leq C  \label{moment10}
\end{equation}%
for some constant $C\in \mathbb{R}$ and all $\Delta \in
\mathcal{B}_{0}(X)$.

Define the set
\begin{equation*}
\hat{\Gamma}_{T}:=\left\{ \hat{\gamma}\in \Gamma (X,S):\phi (\hat{\gamma}%
)\leq T\right\} ,\ T>0,
\end{equation*}%
and observe  that for any set $\Lambda \in
\mathcal{B}_{0}(X)$, there exists a constant $c_{\Lambda }$ such that%
\begin{equation*}
N(\hat{\gamma}_{\Lambda })\leq c_{\Lambda }T,\ \hat{\gamma}\in \hat{\Gamma}%
_{T},\ T>0.
\end{equation*}

Consider the family of measures $\widehat{\Pi }_{m}(d\hat{\gamma}):=\hat{\nu}%
_{\Lambda _{m}}^{\xi }\left( d\hat{\gamma}\right) ,\ m\in
\mathbb{N},$ where $\left\{ \Lambda _{m}\right\} _{m\in
\mathbb{N}}\subset \mathcal{B}_{0}(X)$ is an increasing sequence
exhausting $X$ and $\hat{\nu}_{\Lambda _{m}}^{\xi }(d\hat{\gamma})$
are defined by formula (\ref{nu-measure}).

\begin{prop}
\label{equicont2}The family $\left\{ \widehat{\Pi }_{m}\right\}
_{m\in \mathbb{N}}$ is LEC.
\end{prop}

\noindent \textbf{Proof. }Fix $\Delta \in \mathcal{B}_{0}(X)$ and
$\left\{ B_{n}\right\} _{n\in \mathbb{N}}$ as in Definition
\ref{equi}. It is
sufficient to prove that $\forall \varepsilon >0$ there exist $m_{0}$ and $%
n_{0}$ such that%
\begin{equation*}
\widehat{\Pi }_{m}\left( B_{n}\right) \leq \varepsilon
\end{equation*}%
for any $m\geq m_{0}$ and $n\geq n_{0}$. The following estimate
follows from (\ref{W-est}) by an easy calculation:

\begin{multline*}
E_{\gamma _{\Delta }}(\sigma \left\vert \zeta \right.
)=\sum_{\{x,y\}\subset \gamma _{\Delta }}W_{x,y}(\sigma (x),\sigma
(y))+\sum_{\substack{ x\in \gamma _{\Delta }  \\ y\in \gamma
_{\Delta ^{c}}}}W_{x,y}(\sigma (x),\zeta
(y)) \\
\leq I_{W}\sum_{\{x,y\}\subset \gamma _{\Delta }}\left( \left\vert
\sigma
(x)\right\vert ^{r}+\left\vert \sigma (y)\right\vert ^{r}\right) +J_{W}\frac{%
N(\gamma _{\Delta })\left( N(\gamma _{\Delta })-1\right) }{2} \\
+I_{W}\sum_{\substack{ x\in \gamma _{\Delta }  \\ y\in \gamma
_{\Delta ^{c}}\cap \Delta _{R}}}\left( \left\vert \sigma
(x)\right\vert ^{r}+\left\vert \zeta (y)\right\vert ^{r}\right)
+J_{W}N(\gamma _{\Delta
})N(\gamma _{\Delta ^{c}}\cap \Delta _{R}) \\
\leq I_{W}\left( \left( 2N(\gamma _{\Delta })+N(\gamma _{\Delta
^{c}}\cap \Delta _{R})\right) \sum_{x\in \gamma _{\Delta
}}\left\vert \sigma (x)\right\vert ^{r}+N(\gamma _{\Delta
})\sum_{y\in \gamma _{\Delta ^{c}}\cap
\Delta _{R}}\left\vert \zeta (y)\right\vert ^{r}\right) \\
+J_{W}\left( N(\gamma _{\Delta })^{2}+N(\gamma _{\Delta })N(\gamma
_{\Delta ^{c}}\cap \Delta _{R})\right) .
\end{multline*}%
First, we fix $T>0$ and estimate the corresponding measures of the
sets $B_{n}\cap \hat{\Gamma}_{T}$ and $B_{n}\cap \left(
\hat{\Gamma}_{T}\right) ^{c}$. Taking into account that $\left\vert
u\right\vert ^{r}\leq \left\vert u\right\vert ^{p}+1$for $r<p$, we
obtain the bounds $\sum_{x\in \gamma
_{\Delta }}\left\vert \sigma (x)\right\vert ^{r}\leq \phi (\hat{\gamma}%
)+N(\gamma _{\Delta })$, $\sum_{y\in \gamma _{\Delta ^{c}}\cap
\Delta _{R}}\left\vert \zeta (y)\right\vert ^{r}\leq \phi
(\hat{\gamma})+N(\gamma _{\Delta ^{c}}\cap \Delta _{R})$, where
$\hat{\gamma}=\left( \gamma ,\sigma _{\gamma _{\Delta }}\times \zeta
_{_{\gamma _{\Delta ^{c}}}}\right) $. Thus for $\hat{\gamma}\in
\hat{\Gamma}_{T}$ we have $\sum_{x\in \gamma _{\Delta }}\left\vert
\sigma (x)\right\vert ^{r}\leq cT$, $\sum_{y\in \gamma _{\Delta
^{c}}\cap \Delta _{R}}\left\vert \zeta (y)\right\vert ^{r}\leq cT$
for some constant $c$, which in turn implies that
\begin{equation*}
\mathbf{1}_{B_{n}\cap \hat{\Gamma}_{T}}\left( \hat{\gamma}\right)
\left\vert E_{\gamma _{\Delta }}(\sigma \left\vert \zeta \right.
)\right\vert \leq I_{W}\left( 3cT^{2}+cT^{2}\right) +2T^{2}J_{W}.
\end{equation*}%
Thus there exists a constant $a(T)$ such that%
\begin{equation*}
\mathbf{1}_{B_{n}\cap \hat{\Gamma}_{T}}\left( \hat{\gamma}\right) \mathrm{exp%
}\left\{ -E_{\gamma _{\Delta }}(\sigma \left\vert \zeta \right.
)\right\} \leq a(T)
\end{equation*}%
and
\begin{eqnarray*}
\mathbf{1}_{B_{n}\cap \hat{\Gamma}_{T}}\left( \hat{\gamma}\right)
Z_{\gamma _{\Delta }}^{-1}(\zeta ) &\leq &\mathrm{exp}\left\{
\int_{S^{\gamma _{\Delta }}}\mathbf{1}_{B_{n}\cap
\hat{\Gamma}_{T}}\left( \hat{\gamma}\right)
E_{\gamma _{\Delta }}(\sigma \left\vert \zeta \right. )\chi _{\gamma _{%
\mathcal{\Delta }}}(d\sigma )\right\} \\
&\leq &a(T)
\end{eqnarray*}%
for all $\hat{\gamma}\in \Gamma (X,S)$ and $n\in \mathbb{N}$.

By Chebyshev's inequality applied to measure $\widehat{\Pi }_{m}$
on $\Gamma (X,S)$ we have%
\begin{equation*}
\widehat{\Pi }_{m}\left( \left\{ \hat{\gamma}\in \Gamma (X,S):\phi
\left( \hat{\gamma}\right) \geq T\right\} \right) \leq
T^{-2}\int_{\Gamma (X,S)}\left\vert \phi (\hat{\gamma})\right\vert
^{2}\widehat{\Pi }_{m}\left( d\hat{\gamma}\right)
\end{equation*}%
holding for each $T>0$, which together with (\ref{moment10}) shows
yields
\begin{equation}
\widehat{\Pi }_{m}\left( \left( \hat{\Gamma}_{T}\right) ^{c}\right)
\leq \varepsilon,  \label{cheb}
\end{equation}%
holding for each $\varepsilon >0$ and $T$ bigger than some $T(\varepsilon ,\hat{\zeta%
})$. On the other hand,
\begin{equation*}
\widehat{\Pi }_{m}\left( B_{n}\cap \hat{\Gamma}_{T}\right)
=\int_{\Gamma (X,S)}\mathbf{1}_{B_{n}\cap \hat{\Gamma}_{T}}\left(
\hat{\gamma}\right) \widehat{\Pi }_{m}\left( d\hat{\gamma}\right)
=\int_{\Gamma (X)}I(\gamma )\mu (d\gamma ),
\end{equation*}%
where%
\begin{equation*}
I(\gamma ):=\int_{S^{\gamma }}\mathbf{1}_{B_{n}\cap
\hat{\Gamma}_{T}}\left( \gamma ,\sigma \right) \Pi _{\gamma
_{\Lambda _{m}}}(d\sigma \left\vert \xi _{\gamma }\right. ).
\end{equation*}%
Observe that there exists $m_{0}$ such that $\Lambda _{m}\supset
\Delta $
for $m\geq m_{0}$. For all such $m$, it follows from consistency property (%
\ref{specification}) that%
\begin{multline*}
I(\gamma )=\int_{S^{\gamma }}\left[ \int_{S^{\gamma
}}\mathbf{1}_{B_{n}\cap \hat{\Gamma}_{T}}\left( \gamma ,\sigma
\right) \Pi _{\gamma _{\Delta }}\left( d\sigma \left\vert \zeta
\right. \right) \right] \Pi _{\gamma
_{\Lambda _{m}}}(d\zeta \left\vert \xi _{\gamma }\right. ) \\
=\int_{S^{\gamma }}\mathbf{1}_{B_{n}\cap \hat{\Gamma}_{T}}\left(
\gamma
,\zeta \right) Z_{\gamma _{\Delta }}^{-1}(\zeta ) \\
\times \left[ \int_{S^{\gamma }}\mathbf{1}_{B_{n}\cap \hat{\Gamma}%
_{T}}\left( \gamma ,\sigma _{\gamma _{\Delta }}\times \zeta
_{_{\gamma _{\Delta ^{c}}}}\right) ~\mathrm{exp}~\left( -E_{\gamma
_{_{\Delta }}}(\sigma \left\vert \zeta \right. )\right) \chi
_{\gamma _{\Delta }}(d\sigma )\right] \Pi _{\gamma _{\Lambda
_{m}}}(d\zeta \left\vert \xi
_{\gamma }\right. ) \\
\leq a(\Delta ,T)^{2}\int_{S^{\gamma }}\mathbf{1}_{B_{n}\cap \hat{\Gamma}%
_{T}}\left( \gamma ,\sigma _{\gamma _{\Delta }}\times \xi _{_{\gamma
_{\Delta ^{c}}}}\right) \chi _{\gamma _{\Delta }}(d\sigma ).
\end{multline*}%
Therefore
\begin{multline*}
\int_{\Gamma (X,S)}\mathbf{1}_{B_{n}\cap \hat{\Gamma}_{T}}\left( \hat{\gamma}%
\right) \widehat{\Pi }_{m}\left( d\hat{\gamma}\right) \leq a(\Delta
,T)^{2}\int_{\Gamma (X)}\int_{S^{\gamma }}\mathbf{1}_{B_{n}\cap \hat{\Gamma}%
_{T}}\left( \gamma ,\sigma _{\gamma _{\Delta }}\times \xi _{_{\gamma
_{\Delta ^{c}}}}\right) \\
\chi _{\gamma _{\Delta }}(d\sigma )\mu (d\gamma )<\varepsilon
\end{multline*}%
for $n$ greater than some $n(\varepsilon ,T)$ since
$B_{n}\rightarrow
\emptyset ,\ n\rightarrow \infty $. Combining this with estimate (\ref{cheb}%
) we can see that $\forall \varepsilon >0$ and $m\geq m_{0}$, $n\geq
n_{0}=n(\varepsilon /2,T(\varepsilon /2))$ we have%
\begin{multline*}
\widehat{\Pi }_{m}\left( B_{n}\right) =\widehat{\Pi }_{m}\left(
B_{n}\cap \left( \hat{\Gamma}_{T}\right) ^{c}\right) +\widehat{\Pi
}_{m}\left(
B_{n}\cap \hat{\Gamma}_{T}\right) \\
\leq \widehat{\Pi }_{m}\left( \left( \hat{\Gamma}_{T}\right)
^{c}\right) +\int_{\Gamma _{\Delta }(,S)}\mathbf{1}_{B_{n}\cap
\hat{\Gamma}_{T}}\left( \hat{\gamma}_{\Delta }\right) \widehat{\Pi
}_{m}\left( d\hat{\gamma}\right) \leq \varepsilon /2+\varepsilon
/2=\varepsilon .
\end{multline*}%
The proof is complete.
\hfill%
$\square $

\begin{corollary}
The family of measures $\left\{ \hat{\nu}_{\Delta }^{\xi },\Delta
\in \mathcal{B}_{0}(X)\right\} $ contains a sequence
$\hat{\nu}_{\Delta _{n}}^{\xi },\ n\in \mathbb{N}$, which $\tau
_{loc}$-converges to a probability measure $\hat{\nu}^{\xi }$ on
$\Gamma (X,S)$. Without loss of generality we can assume that the
sequence of sets $\Delta _{n}$ is increasing and exhausts $X$.
\end{corollary}

Let $\nu _{X}\ $be the projection of the measure $\hat{\nu}^{\xi }$ onto $%
\Gamma (X)$ and $F\in L_{0}^{\infty }(\Gamma (X))$. For the function $\hat{F}%
:=F\circ p_{X}\in L_{0}^{\infty }(\Gamma (X,S))$ we have%
\begin{equation*}
\int_{\Gamma (X,S)}\hat{F}(\hat{\gamma})\hat{\nu}_{\Delta _{n}}^{\xi }(d\hat{%
\gamma})\rightarrow \int_{\Gamma (X)}F(\gamma )\nu _{X}(d\gamma ),\
n\rightarrow \infty .
\end{equation*}%
On the other hand, taking the limit in
\begin{equation*}
\int_{\Gamma (X,S)}\hat{F}(\hat{\gamma})\hat{\nu}_{\Delta _{n}}^{\xi }(d\hat{%
\gamma})=\int_{\Gamma (X)}F(\gamma )\left( \int_{S^{\gamma }}\Pi
_{\gamma _{\Delta _{n}}}(d\sigma \left\vert \xi _{\gamma }\right.
)\right) \ \mu (d\gamma )=\int_{\Gamma (X)}F(\gamma )\mu (d\gamma )
\end{equation*}%
we get
\begin{equation*}
\int_{\Gamma (X,S)}\hat{F}(\hat{\gamma})\hat{\nu}^{\xi }(d\hat{\gamma}%
)=\int_{\Gamma (X)}F(\gamma )\mu (d\gamma ),
\end{equation*}%
which in turn implies
\begin{equation*}
\mu =\nu _{X}.
\end{equation*}%
The application of Theorem 8.1 of \cite{Par} to the measurable map $%
p_{X}:\Gamma (X,S)\rightarrow \Gamma (X)$ yields the existence of
the corresponding regular conditional probability distribution $\nu
_{\gamma }^{\xi }$, $\gamma \in \Gamma (X)$, that is, a family of
probability measures $\nu _{\gamma }^{\xi }$ on $\Gamma (X,S)$ such
that for any measurable set $A\subset \Gamma (X,S)$, it follows that
\begin{equation*}
\hat{\nu}^{\xi }(A)=\int_{\Gamma (X)}\nu _{\gamma }^{\xi }(A)\mu
(d\gamma ),
\end{equation*}%
and the map%
\begin{equation}
\Gamma (X)\ni \gamma \mapsto \nu _{\gamma }^{\xi }(A)
\label{f-meas1}
\end{equation}%
is measurable. Moreover, $\nu _{\gamma }^{\xi }[\Gamma
(X,S)\diagdown p_{X}(\gamma )]=0$ for $\mu $-a.a. $\gamma \in \Gamma
(X)$. Thus $\nu _{\gamma }^{\xi }$ generates (for $\mu $-a.a.
$\gamma \in \Gamma (X)$) a probability measure on $p_{X}(\gamma
)=S^{\gamma }$ (for which we preserve
the notation $\nu _{\gamma }^{\xi }$) such that the map%
\begin{equation*}
\Gamma (X):\gamma \mapsto \int_{S^{\gamma }}F(\gamma ,\sigma )\nu
_{\gamma }^{\xi }(d\sigma )
\end{equation*}%
is measurable for any $F\in L_{0}^{\infty }(\Gamma (X,S))$ and%
\begin{equation}
\int_{\Gamma (X,S)}F(\hat{\gamma})\hat{\nu}^{\xi }(d\hat{\gamma}%
)=\int_{\Gamma (X)}\left( \int_{S^{\gamma }}F(\gamma ,\sigma )\nu
_{\gamma }^{\xi }(d\sigma )\right) \mu (d\gamma ).
\label{desintegr}
\end{equation}

\begin{prop}
\label{th-gibbs copy(1)}For $\mu $-a.a. $\gamma \in \Gamma (X)$ and
any $\xi $ of the form (\ref{section}), we have that
\begin{equation*}
\nu _{\gamma }^{\xi }\in \mathcal{G}_{\alpha ,p}(S^{\gamma }).
\end{equation*}
\end{prop}

\noindent \textbf{Proof. }We first prove that $\nu _{\gamma }^{\xi
}$
satisfies the DLR equation (\ref{DLR}). Fix $\Lambda \subset \mathcal{B}_{0}%
\mathbf{(}X)$ and $\Delta _{n}$ such that $\Lambda \subset \Delta
_{n}$. According to consistency property (\ref{specification}) we
have
\begin{equation*}
\int_{S^{\gamma }}\Pi _{\gamma _{\Lambda }}(d\sigma \left\vert \zeta
\right. )\Pi _{\gamma _{\Delta _{n}}}(d\zeta \left\vert \xi \right.
)=\Pi _{\gamma _{\Delta _{n}}}(d\sigma \left\vert \xi \right. ),
\end{equation*}%
so that%
\begin{equation*}
\int_{\Gamma (X,S)}g(\gamma ,\zeta )\Pi _{\gamma _{\Delta
_{n}}}(d\zeta \left\vert \xi _{\gamma }\right. )\mu (d\gamma
)=\int_{\Gamma (X,S)}G(\gamma ,\sigma )\Pi _{\gamma _{\Delta
_{n}}}(d\sigma \left\vert \xi _{\gamma }\right. )\mu (d\gamma ),
\end{equation*}%
where $G\in L_{0}^{\infty }(\Gamma (X,S))$, $g(\gamma ,\zeta
):=\int_{S^{\gamma }}G(\gamma ,\sigma )\Pi _{\gamma _{\Lambda
}}(d\sigma
\left\vert \zeta \right. )$ and $\xi _{\gamma }$ is given by (\ref{section}%
). Observe that $g\in L_{0}^{\infty }(\Gamma (X,S)),$ so that we can
pass to
the limit as $n\rightarrow \infty $ and obtain the equality%
\begin{equation*}
\int_{\Gamma (X,S)}g(\hat{\gamma})\hat{\nu}^{\xi }(d\hat{\gamma}%
)=\int_{\Gamma (X,S)}G(\hat{\gamma})\hat{\nu}^{\xi }(d\hat{\gamma}),
\end{equation*}%
or%
\begin{equation*}
\int_{\Gamma (X)}\int_{S^{\gamma }}\int_{S^{\gamma }}G(\gamma
,\sigma )\Pi _{\gamma _{\Lambda }}(d\sigma \left\vert \zeta \right.
)\nu _{\gamma }^{\xi }(d\zeta )\mu (d\gamma )=\int_{\Gamma
(X)}\int_{S^{\gamma }}G(\gamma ,\sigma )\nu _{\gamma }^{\xi
}(d\sigma )\mu (d\gamma ),
\end{equation*}%
which in turn implies that%
\begin{equation*}
\int_{S^{\gamma }}\Pi _{\gamma _{\Lambda }}(d\sigma \left\vert \zeta
\right. )\nu _{\gamma }^{\xi }(d\zeta )=\nu _{\gamma }^{\xi
}(d\sigma )
\end{equation*}%
for a.a. $\gamma $. Thus, (\ref{DLR}) does hold.

In order to prove that $\nu _{\gamma }^{\xi }$ is supported on
$l_{\alpha
}^{p}(\gamma ,S)$ for a.a. $\gamma \in \Gamma (X)$, we introduce the cut-off%
\begin{equation*}
\phi _{L,K}(\hat{\gamma}):=\sum_{\left\vert k\right\vert \leq
K}e^{-\alpha \left\vert k\right\vert }\left( N(\gamma _{k})\wedge
L\right) +\left( \left\Vert \sigma _{\gamma }\right\Vert _{\alpha
,p}^{p}\wedge L\right) ,
\end{equation*}%
$K,L\in \mathbb{N}$, and observe that
\begin{equation*}
\int_{\Gamma (X,S)}\phi (\hat{\gamma})\hat{\nu}_{\Delta _{n}}^{\xi }(d\hat{%
\gamma})=\lim_{K\rightarrow \infty }\lim_{L\rightarrow \infty
}\int_{\Gamma (X,S)}\phi _{L,K}(\hat{\gamma})\hat{\nu}_{\Delta
_{n}}^{\xi }(d\hat{\gamma}).
\end{equation*}%
Moreover, $\phi _{L,K}\in L_{0}^{\infty }(\Gamma (X,S))$, and the
limit
transition as $n\rightarrow \infty $ together with the estimate (\ref%
{moment10}) show that%
\begin{equation*}
\int_{\Gamma (X,S)}\phi (\hat{\gamma})\hat{\nu}^{\xi
}(d\hat{\gamma})<\infty .
\end{equation*}%
Thus%
\begin{equation*}
\int_{\Gamma (X)}\int_{S^{\gamma }}\left\Vert \sigma _{\gamma
}\right\Vert _{\alpha ,p}^{p}\nu _{\gamma }^{\xi }(d\sigma _{\gamma
})\mu (d\gamma )\leq \int_{\Gamma (X,S)}\phi
(\hat{\gamma})\hat{\nu}^{\xi }(d\hat{\gamma})<\infty ,
\end{equation*}%
so that $\int_{S^{\gamma }}\left\Vert \sigma _{\gamma }\right\Vert
_{\alpha ,p}^{p}\nu _{\gamma }^{\xi }(d\sigma _{\gamma })<\infty $
for a.a. $\gamma \in \Gamma (X)$, which in turn implies that
\newline $\nu _{\gamma }^{\xi }(l_{\alpha }^{p}(\gamma ,S))=1$.
\hfill $\square $

\noindent\textbf{Proof of Theorem \ref{th-meas}. }The result follows
directly from formula (\ref{f-meas1}) and Proposition \ref{th-gibbs
copy(1)}. \hfill $\square $

\begin{rem}
Let $\nu _{\gamma }\in \mathcal{G}_{\alpha ,p}(S^{\gamma })$,
$\gamma \in \Gamma (X),$ be a family of Gibbs measures satisfying
the measurability
condition (\ref{meas}). For $\mu $-a.a. $\gamma \in \Gamma (X)$ the measure $%
\nu _{\gamma }$ obeys the moment estimate (\ref{moment-est}).
Integrating
both sides of this inequality we obtain%
\begin{multline*}
\int_{\Gamma (X)}\left( \int_{S^{\gamma }}\left\Vert \sigma
\right\Vert _{\alpha ,p}^{p}~\nu _{\gamma }(d\sigma )\right) \mu
(d\gamma )\leq
C_{1}\int_{\Gamma (X)}b_{\alpha }(\gamma )\mu (d\gamma ) \\
+C_{2}\int_{\Gamma (X)}a_{\alpha ,p^{\prime }}\left( \gamma \right)
\mu (d\gamma )<\infty
\end{multline*}%
because of Proposition \ref{prop1}.
\end{rem}

Let us note that the convergence of the measures $\hat{\nu}_{\Delta
_{n}}^{\xi },$ $n\in \mathbb{N}$, to $\hat{\nu}^{\xi }$ does not in
general imply the convergence of their conditional distributions
$\Pi _{\gamma _{\Delta _{n}}}(d\sigma \left\vert \xi _{\gamma
}\right. )$ to $\nu _{\gamma }^{\xi
}(d\sigma )$ for $\mu $-a.a. $\gamma $. However, we can make use of Koml\'{o}%
s' theorem (see e.g. \cite{Bal}\textbf{) }and\textbf{\ }prove the
following result.

\begin{prop}
There exists a sequence $(n_{j},\ j\in \mathbb{N})\subset
\mathbb{N}$ such that for $\mu $-a.a. $\gamma \in \Gamma (X)$ and
$\gamma _{j}:=\gamma _{\Delta _{n_{j}}}$ we have the local setwise
convergence of measures
\begin{equation*}
\frac{1}{N}\sum_{j=1}^{N}\Pi _{\gamma _{j}}(\cdot \left\vert \xi
_{\gamma }\right. )\rightarrow \nu _{\gamma }^{\xi },\ \
N\rightarrow \infty .
\end{equation*}
\end{prop}

\noindent\textbf{Proof. }It has been shown in \cite{DKKP} that there
exists a countable family of functions $\left\{ f_{m},\ m\in
\mathbb{N}\right\}
\subset L_{0}^{\infty }(\Gamma (X,S))$, which form a separating class for $%
\mathcal{P}(\Gamma (X,S))$. That is, for any two measures $\mu ,\nu \in\mathcal{P}%
(\Gamma (X,S))$ the equality $\int f_{m}d\mu =\int f_{m}d\nu $,
$m\in \mathbb{N}$,  implies that $\mu =\nu $. Consider the family of
functions
\begin{equation*}
g_{\Delta _{n}}^{(m)}(\gamma ):=\int_{S^{\mathbb{\gamma
}}}f_{m}(\gamma ,\sigma )\Pi _{\gamma _{\Delta _{n}}}(d\sigma
\left\vert \xi _{\gamma }\right. ),\ \ \ n,m\in \mathbb{N}.
\end{equation*}%
It is clear that $g_{\Delta _{n}}^{(m)}\in L^{1}(\Gamma (X),\mu )$.
Applying Koml\'{o}s' theorem and the diagonal procedure (as in the
proof of
Theorem 3.6 in \cite{KKP2}) one can show that there exists a sequence $%
(n_{j},\ j\in \mathbb{N})\subset \mathbb{N}$ such that for  $\mu $-a.a. $%
\gamma \in \Gamma (X)$, all $m\in \mathbb{N}$ and any subsequence $%
(n_{j_{k}},\ k\in \mathbb{N})$ we have the Cesaro means convergence%
\begin{equation}
\frac{1}{N}\sum_{k=1}^{N}g_{j_{k}}^{(m)}(\gamma )\rightarrow
g^{(m)}(\gamma ),\ N\rightarrow \infty ,  \label{ces1}
\end{equation}%
where $g_{j}^{(m)}:=g_{\Delta _{n_{j}}}^{(m)}$ and $g^{(m)}\in
L^{1}(\Gamma (X),\mu )$. Observe that $g^{(m)}$ is independent of
the chioce of the subsequence $(n_{j_{k}},\ k\in \mathbb{N})$.

Moreover, for  $\mu $-a.a. $\gamma \in \Gamma (X)$ the family of measures $%
\Pi _{\gamma _{j}}(d\sigma \left\vert \xi _{\gamma }\right. ):=\Pi
_{\gamma _{\Delta _{n_{j}}}}(d\sigma \left\vert \xi _{\gamma
}\right. )$, $j\in \mathbb{N}$, is relatively compact in
$\tau_{loc}$ topology (see proof of Theorem \ref{gibbs1}) and thus
contains a sequence $\Pi _{\gamma _{j_{k}}}(d\sigma \left\vert \xi
_{\gamma }\right. )$ converging to some measure $\eta _{\gamma
}^{\xi }(d\sigma )$ on $S^{\gamma }$. This together with
(\ref{ces1}) implies  the equality
\begin{equation}
g^{(m)}(\gamma )=\int_{S^{\mathbb{\gamma }}}f_{m}(\gamma ,\sigma
)\eta _{\gamma }^{\xi }(d\sigma )  \label{g1}
\end{equation}%
for all $m\in \mathbb{N}$ and  $\mu $-a.a. $\gamma \in \Gamma (X)$.
Integration of both sides of (\ref{g1}) shows that%
\begin{equation}
\int_{\Gamma (X)}g^{(m)}(\gamma )\mu (d\gamma )=\int_{\Gamma
(X,S)}f_{m}(\gamma ,\sigma )\eta _{\gamma }^{\xi }(d\sigma )\mu
(d\gamma ). \label{g11}
\end{equation}%
On the other hand, the convergence $\hat{\nu}_{\Delta _{n}}^{\xi
}\rightarrow $ $\hat{\nu}^{\xi }$ together with (\ref{desintegr})
implies
that%
\begin{equation*}
\int_{\Gamma (X,S)}f_{m}(\gamma ,\sigma
)\frac{1}{N}\sum_{j=1}^{N}\Pi _{\gamma _{j}}(d\sigma \left\vert \xi
_{\gamma }\right. )\mu (d\gamma )\rightarrow \int_{\Gamma
(X,S)}f_{m}(\gamma ,\sigma )\nu _{\gamma }^{\xi }(d\sigma )\mu
(d\gamma ),
\end{equation*}%
$\ N\rightarrow \infty ,$ so that%
\begin{equation}
\int_{\Gamma (X)}g^{(m)}(\gamma )\mu (d\gamma )=\int_{\Gamma
(X,S)}f_{m}(\gamma ,\sigma )\nu _{\gamma }^{\xi }(d\sigma )\mu
(d\gamma ) \label{g2}
\end{equation}%
for all $m\in \mathbb{N}$ and $\mu $-a.a. $\gamma \in \Gamma (X)$.
The combination of equalities (\ref{g11}) and (\ref{g2}) together
with the
measure separating property of the family $\left\{ f_{m},\ m\in \mathbb{N}%
\right\} $ shows that $\nu _{\gamma }^{\xi }=\eta _{\gamma }$ for $\mu $%
-a.a. $\gamma \in \Gamma (X)$, which completes the proof.
\hfill$\square $

\section{Acknowledgments}

Financial support by the DFG through SFB 701 \textquotedblleft
Spektrale Strukturen und Topologische Methoden in der
Mathematik\textquotedblright , IRTG 1132 \textquotedblleft
Stochastics and Real World Models\textquotedblright\ and the
German-Polish research project 436 POL 125/0-1, and by the ZiF
Research Group "Stochastic Dynamics: Mathematical Theory and
Applications" (University Bielefeld) is gratefully acknowledged.


\begin{thebibliography}{99}
\bibitem{AKLU} S. Albeverio, Yu. Kondratiev, E. Lytvynov, G. Us, Analysis
and geometry on marked configuration spaces. Infinite dimensional
harmonic analysis (Kyoto, 1999), 1--39, Gr\"{a}bner, Altendorf,
2000.

\bibitem{Bal} E. J. Balder, Infinite-dimensional extensions of a theorem of
Koml\'{o}s, Prob. Th. Rel. Fields 81 (1989) 185--188.


\bibitem{BoDal} L. Bogachev, A. Daletskii. Cluster point processes on
manifolds. J. Geom. Phys. 63 (2013) 45--79.

\bibitem{Bov} A. Bovier, Statistical Mechanics of Disordered Systems. A
Mathematical Perspective. Cambridge Series in Statistical and
Probabilistic Mathematics. Cambridge University Press, Cambridge,
2006.

\bibitem{CG} F. Conrad, M. Grothaus, $N/V$-limit for Langevin dynamics in
continuum, Reviews in Math. Physics 23 (2011), 1--51.

\bibitem{DKKP} A. Daletskii, Yu.~Kondratiev, Yu. Kozitsky, T. Pasurek, Phase
Transitions in a quenched amorphous ferromagnet, Preprint 12142,
SFB\ 701, Universit\"{a}t Bielefeld (2012).

\bibitem{DVJ1} D.J. Daley, D. Vere-Jones, An Introduction to the Theory of
Point Processes Volume I: Elementary Theory and Methods,\ 2nd
edition (Springer, New York, 2003).

\bibitem{Do70} R. L. Dobrushin, Prescribing a system of random variables by
conditional distributions, Theory Probab. Appl. 15 (1970), 101--118.

\bibitem{Geor} H.-O. Georgii, Gibbs Measures and Phase Transitions, De
Gruyter Studies in Mathematics Vol. 9, Berlin: de Gruyter, 1988.

\bibitem{GHM} H.-O. Georgii, O. H\"{a}ggstr\"{o}m, C. Maes, The random
geometry of equilibrium phases. In: C. Domb and J.L. Lebowitz (eds.)
Phase Transitions and Critical Phenomena Vol. 18, Academic Press,
London 2000, pp. 1-142.

\bibitem{Kal} O. Kallenberg, Random Measures, 3rd edition, Berlin,
Akademie-Verlag, 1983.

\bibitem{KMM} J. Kerstan, K. Matthes, J. Mecke, Infinitely Divisible Point
Processes, Wiley \& Sons, 1978.

\bibitem{KKP1} Yu.~Kondratiev, Yu. Kozitsky, T. Pasurek, Gibbs random fields
with unbounded spins on unbounded degree graphs, J. Appl. Probab.\
47 (2010), 856-875.

\bibitem{KKP2} Yu.~Kondratiev, Yu. Kozitsky, T. Pasurek, Gibbs measures of
disordered lattice systems with unbounded spins, Markov Processes
Relat. Fields 18 (2012), 553--582.

\bibitem{KKun} Yu. G. Kondratiev, T. Kuna, Harmonic analysis on
configuration space I. General theory, Infin. Dimens. Anal. Quantum
Probab. Relat. Top.\ 5 (2002), 201--233.

\bibitem{KKut} Yu.~Kondratiev, O. Kutovyi, On the metrical properties of the
configuration space, Math. Nachr. 279, No. 7, 774--783 (2006).

\bibitem{KPR} Yu.~Kondratiev, T. Pasurek, M. R\"{o}ckner, Gibbs measures of
continuous systems: an analytic approach, Reviews in Math. Physics
24 (2012).

\bibitem{KP} Yu. Kozitsky, T. Pasurek, Euclidean Gibbs measures of
interacting quantum anharmonic oscillators, J. Stat. Phys. 127
(2007), 985--1047.

\bibitem{KunaPhD} T.~Kuna, Studies in Configuration Space Analysis and
Applications, Ph.D.\ dissertation, Rheinische Friedrich-Wilhelms-Universit%
\"{a}t Bonn, 1999, in: Bonner Math.\ Schrift. 324, Universit\"{a}t
Bonn, Math.\ Inst., Bonn, 1999, 187~pp.


\bibitem{LP} J. L. Lebowitz, E. Presutti, Statistical mechanics of systems
of unbounded spins, Commun. Math. Phys. 50 (1976), 195--218.

\bibitem{Len73} A. Lenard, Correlation functions and the uniqueness of the
state in classical statistical mechanics. Commun. Math. Phys. 30
(1973), 35--44.



\bibitem{New} C. M.\ Newman, Topics in disordered systems. Lectures in
Mathematics ETH Z\"{u}rich, Birkh\"{a}ser Verlag, Basel, 1997.

\bibitem{NS} C. M.\ Newman, D. L. Stein, Thermodynamic chaos and the
structure of the short-range spin glasses. In Mathematical aspects
of spin glasses and neural networks, eds. A. Bovier and P. Picco,
243-287, Progr. Probab.,41, Birkh\"{a}ser Boston, Boston MA, 1998.

\bibitem{Par} K.R.~Parthasarathy, Probability Measures on Metric Spaces,
Probab.\ Math.\ Statist., Academic Press, New York, 1967.

\bibitem{Pre} Ch. Preston, Random Fields, Lect. Notes Math.\ 534 (Springer,
Berlin, 1976).

\bibitem{Res} S. Resnick, Extreme Values, Regular Variation, and Point
Processes, Applied Probability, Springer, New York, 1987.

\bibitem{Ru69} D. Ruelle, Statistical Mechanics. Rigorous Results
(Benjamins, New York, 1969).

\bibitem{Ru70} D. Ruelle, Superstable interactions in classical statistical
mechanics, Commun. Math. Phys. 18 (1970), 127--159.
\end{thebibliography}
\end{document}